\documentclass[prd,aps,showpacs,nofootinbib,eqsecnum,amsfonts,amsmath,twocolumn,floatfix,a4paper]{revtex4}
\usepackage{epsfig}
\usepackage{graphics,color}
\usepackage{graphicx}
\usepackage{bm}
\usepackage{calc}
\usepackage{psfrag}

\begin{document}


\newcommand{\geo}{\textsc{GEO\,600}}
\newcommand{\bth}{\bf \tilde h}
\newcommand{\btx}{\bf \tilde x}
\newcommand{\x}{$X$}
\newcommand{\h}{$H$}
\newcommand{\ie}{i.e.}

\newcommand{\micep}{$F_{\rm ref}$}
\newcommand{\micvis}{$P_{\rm ref}$}
\newcommand{\miphase}{$\Phi_{\rm MI}$}
\newcommand{\miamp}{$A_{\rm MI}$}

\newcommand{\Ex}{{\bf E}_{\rm X}}
\newcommand{\dEx}{\Delta {\bf E}_{\rm X}}
\newcommand{\dEh}{\Delta {\bf E}_{\rm H}}
\newcommand{\dE}{\Delta {\bf E}}
\newcommand{\Exa}{E^{a}_{\rm X}}
\newcommand{\Exf}{E^{f}_{\rm X}}
\newcommand{\Ext}{E^{t}_{\rm X}}
\newcommand{\Exb}{E^{b}_{\rm X}}
\newcommand{\Eh}{{\bf E}_{\rm H}}
\newcommand{\Eha}{E^{a}_{\rm H}}
\newcommand{\Ehf}{E^{f}_{\rm H}}
\newcommand{\Eht}{E^{t}_{\rm H}}
\newcommand{\df}{\mathrm{d}f}
\newcommand{\dt}{\mathrm{d}t}
\newcommand{\dx}{\mathrm{d}x}
\newcommand{\dwf}{\mathrm{d}w^f}
\newcommand{\dwt}{\mathrm{d}w^t}
\newcommand{\dwa}{\mathrm{d}w^a}
\newcommand{\Txhf}{T_\mathrm{XH}(f)}
\newcommand{\Sx}{{\bf S}_{\rm X}}
\newcommand{\Sh}{{\bf S}_{\rm H}}
\newcommand{\bH}{\bf h}
\newcommand{\bX}{\bf x}
\newcommand{\tX}{\tilde x}
\newcommand{\tH}{\tilde h}
\newcommand{\btdelta}{\bm{\tilde \delta}}
\newcommand{\etal}{{\it et al}}
\newcommand{\be}{\begin{equation}}
\newcommand{\ee}{\end{equation}}
\newcommand{\ber}{\begin{eqnarray}}
\newcommand{\eer}{\end{eqnarray}}

\newenvironment{Ventry}[1]%
{\begin{list}{}%
    {\renewcommand{\makelabel}[1]{\textbf{##1:}\hfil}%
     \settowidth{\labelwidth}{\textbf{#1:}}%
     \setlength{\leftmargin}{\labelwidth+\labelsep}%
    }%
}
{\end{list}}

\title {Physical instrumental vetoes for gravitational-wave burst triggers}

\author{P. Ajith}\email{Ajith.Parameswaran@aei.mpg.de}
\affiliation{Max-Planck-Institut f\"ur Gravitationsphysik 
(Albert-Einstein-Institut) and Leibniz Universit\"at Hannover,
Callinstr.~38, 30167 Hannover, Germany}

\author{M. Hewitson}
\affiliation{Max-Planck-Institut f\"ur Gravitationsphysik
(Albert-Einstein-Institut) and Leibniz Universit\"at Hannover,
Callinstr.~38, 30167 Hannover, Germany}

\author{J. R. Smith}
\affiliation{Max-Planck-Institut f\"ur Gravitationsphysik
(Albert-Einstein-Institut) and Leibniz Universit\"at Hannover,
Callinstr.~38, 30167 Hannover, Germany}

\author{H. Grote}
\affiliation{Max-Planck-Institut f\"ur Gravitationsphysik
(Albert-Einstein-Institut) and Leibniz Universit\"at Hannover,
Callinstr.~38, 30167 Hannover, Germany}

\author{S. Hild}
\affiliation{Max-Planck-Institut f\"ur Gravitationsphysik
(Albert-Einstein-Institut) and Leibniz Universit\"at Hannover,
Callinstr.~38, 30167 Hannover, Germany}

\author{K. A. Strain}
\affiliation{Department of Physics and Astronomy, 
University of Glasgow, Glasgow, G12 8QQ, United Kingdom}
\affiliation{Max-Planck-Institut f\"ur Gravitationsphysik
(Albert-Einstein-Institut) and Leibniz Universit\"at Hannover,
Callinstr.~38, 30167 Hannover, Germany}

\date{\today}


\begin{abstract}

We present a robust strategy to \emph{veto} certain classes of instrumental 
glitches that appear at the output of interferometric gravitational-wave (GW) detectors.
This veto method is `physical' in the sense that, in order to veto a burst trigger, 
we make use of our knowledge of the coupling of different detector subsystems to the main 
detector output. The main idea behind this method is that the noise in an instrumental 
channel \x\ can be \emph{transferred} to the detector output (channel \h) using the 
\emph{transfer function} from \x\ to \h, provided the noise coupling is \emph{linear} and the 
transfer function is \emph{unique}. If a non-stationarity in channel \h\ is causally 
related to one in channel \x, the two have to be consistent with the transfer function. 
We formulate two methods for testing the consistency between the burst triggers in 
channel \x\ and channel \h. One method makes use of the \emph{null-stream} constructed from 
channel \h\ and the \emph{transferred} channel \x, and the second involves cross-correlating 
the two. We demonstrate the efficiency of the veto by `injecting' instrumental glitches
in the hardware of the \geo\ detector. The \emph{veto safety} is demonstrated by performing
GW-like hardware injections. We also show an example application of this method using
5 days of data from the fifth science run of \geo. The method is found to have very high 
veto efficiency with a very low accidental veto rate.

\end{abstract}


\maketitle


\section {Introduction}

The existence of gravitational waves (GWs) is the last and most intriguing prediction
of Einstein's General Theory of Relativity (GTR), which is yet to be verified by a {\it direct}
observation. A worldwide network of GW detectors consisting of 
ground-based interferometers~\cite{ligo,virgo,geo,tama} and resonant 
bars~\cite{explNautilus,auriga,allegro}, most of them operating at or close to their design sensitivity, 
has started looking for signatures of GWs produced by astrophysical and cosmological 
sources. Apart from providing excellent tests of GTR in the 
strongly-gravitating regions~\cite{WillLR}, GW astronomy is expected to open a new window
to the universe providing unique information about various astrophysical 
phenomena. For example, GW observations from the final coalescence of neutron 
star binaries will shed light on the nuclear equation of state~\cite{CutlerEtal93}, 
and the GWs produced by a neutron star or small black hole spiraling into a massive
black hole will provide a `map' of the spacetime geometry around the larger object~\cite{Ryan95}.
Also, GW observations can be used to measure the Hubble constant, deceleration parameter,
and cosmological constant~\cite{SchutzHubble,Markovic93,CF93}. Similarly, 
the observation of the stochastic GW background from the early universe can test 
a number of current speculations about the very early universe~\cite{AllenRev96}.  

The best understood (and, perhaps, the most promising) astrophysical sources of GWs 
for ground-based interferometers are the inspiralling compact binaries consisting of 
black holes and/or neutron stars. What makes this class of sources extremely interesting 
is that the expected waveforms can be very accurately modelled (and easily parametrized
by a few intrinsic parameters of the binary, like the component masses and spins) by 
approximation techniques in GTR. This allows the data analysts to use 
the matched filtering technique in order to extract the signal that is buried in 
the detector noise. Another important class of astrophysical sources which can be probed
through optimal filtering technique is spinning neutron stars and pulsars. 

There are other classes of GW sources, like core-collapse of massive stars in supernovae, binary
black hole/neutron star mergers, accretion induced collapse of white dwarf stars, Gamma 
ray bursts etc., for which the physics is largely 
unknown, or too complex to allow computation of detailed gravitational waveforms. These 
kind of sources are generally classified as `unmodelled burst sources'. 
GW observations from merging black hole binaries can potentially bring very useful 
insights to the nonlinear dynamics of the spacetime curvature as the two black holes
convert themselves to form a single black hole. 
GWs from a Gamma ray burst are expected to carry detailed information about its 
source, which can not be probed via electromagnetic observations. 
Similarly, correlated neutrino and GW observations from a core-collapse supernova could 
bring interesting insights into the newborn neutron star/black hole in the core of the 
supernova (See~\cite{CutlerThorne02} for a detailed review). 

Any possible GW signal coming from an astrophysical/cosmological source is generally 
buried in the detector noise. To extract these GW signatures from the noise is a 
nontrivial data analysis problem. Indeed, if time evolution of the GW phase is accurately
known, the optimal filter for searching for this signal buried in the noise is the well known 
matched-filter. But, since matched filtering relies on the prior knowledge of the signal, 
it may not be the best detection strategy in the search for unmodelled, short-lived GW 
bursts. In the next section, we give a brief introduction to the data analysis for 
`unmodelled burst sources'. 

\subsection{The search for transient, unmodelled gravitational-wave bursts}

One class of search methods that is being employed in the burst data analysis is based 
on time-frequency decomposition of detector data. These algorithms construct time-frequency 
maps of the time-series data and look for `time-frequency regions' containing excess 
power which are statistically unlikely to be associated with the background noise 
distribution~\cite{ExPower}. Some of these algorithms are based on clustering the 
`time-frequency pixels' containing excess power, and applying another threshold 
on these clusters of pixels~\cite{tfClusters}. Time-frequency detection algorithms 
using basis functions other than the standard Fourier basis functions are also 
proposed~\cite{waveBurst,chatterjiQ}.   
Another class of burst detection algorithms look for slopes or ridges in the 
time-series data, or in its time-frequency representation \cite{slopeFilter,slopeFilter2,andersonBala99}. 
In general, these methods are claimed to be robust in detecting short-lived signals with minimum 
{\it a priori} information. 

While the optimal filtering technique, along with accurate models of the waveforms, 
enables one to accurately estimate the physical parameters (such as masses and spins)
of the GW source, the time-frequency methods, by construction, are unable to accomplish
this. Instead, these algorithms try to parametrize the underlying gravitational 
waveforms using a set of quantities like the characteristic central frequency, duration, 
bandwidth etc. The detection algorithms, implemented in the data analysis pipelines, 
are usually referred to as {\it event trigger generators} (ETGs). 

Since current interferometric detectors are highly complex instruments, the detector
output typically contains a large number of noise transients, or `glitches', of instrumental 
origin which cause the ETGs to generate spurious triggers.
One of the main challenges in the burst data analysis is to distinguish 
these spurious bursts from actual GW bursts. Since the expected GW signals are unmodelled, it is
practically impossible to distinguish these `false alarms' from actual GW bursts based 
on their signal characteristics. One way of dealing with this 
issue is to require that the triggers be coincident (within a time window) 
in multiple detectors located at different parts of the world. Although this `coincidence 
requirement' reduces the list of candidate triggers by a considerable amount, this does 
not completely cure the problem. While coincident instrumental bursts in multiple 
detectors are highly improbable, long data-taking runs (typically several months 
long) using multiple detectors can produce a large number of random coincidences
(potentially thousands per month~\cite{LSCS4Burst}). 
It is thus very important to develop robust techniques to distinguish between true 
GW bursts and spurious instrumental bursts which are coincident in different detectors -- 
popularly known as \emph{veto} techniques. 

Since a number of environmental and instrumental noise sources can potentially couple
to the main detector output, many such noise sources are continuously recorded along 
with the data from the main detector output. The measurement points for time-series data 
within the detector are referred to as `channels'. One class of veto methods is based on
identifying triggers in the `gravitational-wave channel' (the main detector output) which are coincident with triggers
in an instrumental/environmental noise channel. The `coincidence windows' are chosen such
that the `accidental' (random) coincidence rate between the two channels is limited to an 
acceptable amount. See~\cite{LSCS4Burst, DiCredico05, Hild07, vetoTAMA} for some recent work on such 
`statistical vetoes'. Another class of `physical vetoes' is based on 
our understanding of how a GW should (or, should not) appear in certain channels
\cite{asqVeto, PQMon, hNull}. Moreover, a number of `waveform consistency tests'
between multiple detectors are also employed in the burst searches
\cite{h1h2imp,redunVeto,cohereCadonati,klimenko,Chatterji06}.  

In this paper, we demonstrate a veto strategy which makes use of our understanding 
of the physical coupling of various detector subsystems to the detector output. 
This method is different from the physical veto methods discussed above in the 
sense that, here we use the knowledge of the coupling mechanism involved
in transporting glitches from a particular subsystem to the main detector output. 
The main idea behind this method is that the noise in an instrumental 
channel \x\ can be \emph{transferred} into the detector output (channel \h) using the
\emph{transfer function} from \x\ to \h~\footnote{Throughout this paper, channel \x\ 
refers to the measurement point for time-series data from a detector subsystem/environmental 
noise source \x, and channel \h\ refers to the main detector output (the `GW channel').}.
If a noise transient in channel \h\ is causally 
related to one in channel \x, the two have to be consistent with the transfer function.
The basic idea of this method was formulated in~\cite{NPVeto}, which also
demonstrated this in the presence of stationary Gaussian noise. Here, we 
demonstrate a more general formulation which admits non-Gaussian tails in the 
noise distribution and other `real-life effects' (see Sec.~\ref{sec:realLifeSit}). 
We also propose an alternative statistic to test the consistency of the two triggers
(see Sec.~\ref{sec:crossCorr}). Sec.~\ref{sec:vetoMethod} reviews
the basic ideas of this method. In Sec.~\ref{sec:HWinj}, we demonstrate the method
by performing hardware injections in the \geo\ detector~\cite{geo}. The results of the
veto analysis performed on 5 days of data from the fifth science run of \geo\ are 
discussed in Sec.~\ref{sec:applRealData}. In Sec.~\ref{sec:summary}, we summarise
our main findings.


\section{Vetoes using known instrumental couplings}
\label{sec:vetoMethod}
Let $x_i$ and $h_i$ denote the discretely sampled time-series data recorded in 
channel \x\ (which records the noise from a detector subsystem) and channel \h\ 
(the `gravitational-wave channel'), respectively. We denote 
the corresponding discrete Fourier transforms (DFT) by $\tilde x_k$ and $\tilde h_k$, 
respectively. The input and output of a linear, time-invariant system are related by the 
{\it transfer function} $T_k$ of the system, which is defined as
\be
T^{\rm XH}_k \equiv \frac{P^{\rm XH}_k }{P^{\rm XX}_k}.
\label{eq:transfn}
\ee
Here $P_k^{\rm XH} \equiv \overline{\tX_k \tH_k^*}$ is the cross-power spectral density of $x_i$ 
and $h_i$, and $P_k^{\rm XX} \equiv \overline{\tX_k \tX_k^*}$ is the power spectral density of 
$x_i$, where the `bars' indicate ensemble averages. If the coupling of noise between 
channel \x\ and \h\ is linear and the transfer function is time-invariant, the Fourier 
transform of the noise measured in channel \x\ at any time can be {\it transferred} to 
channel \h, by using the transfer function 
\be
\tilde x_k' = \tilde x_k \, T^{\rm XH}_k.
\label{eq:noisetransfer}
\ee

$\tilde x_k$ and $\tilde h_k$ can be thought of as components of two vectors $\btx$ and $\bth$ defined in two
$N$-dimensional linear vector spaces. In the mathematical sense, Eq.(\ref{eq:noisetransfer}) 
maps $\btx$ to the space of $\bth$. In the physical sense, this means that if a noise transient 
originates in \x, one can predict how it will appear in \h. If there exists a noise transient 
in \h\ at the same time~\footnote{The time-coincidence window should be chosen according to 
the typical time scale of the transients that we are concerned with. We use a time window of 
a few tens of milliseconds since the current searches for GW bursts seek to detect bursts of duration $<<$ 1 sec.}, 
we can compare it with the above `prediction'. If a noise transient 
in the channel \h\ is causally related to one in channel \x, the data vectors 
$\btx$ and $\bth$ have to be consistent with the transfer function. This allows us to formulate 
a powerful strategy to veto noise transients originating within the detector. 

The basic idea is the following: firstly, we identify time-coincident burst triggers in channels \x\ 
and \h. We compute the DFTs of two short segments of data in channels \x\ and \h. The length of 
these segments (typically a few tens of milliseconds) is chosen so as to encompass only
the noise transient under investigation. 
If these two noise vectors are \emph{consistent} with the transfer function, as given by Eq.(\ref{eq:transfn}), 
it is highly likely that the noise transient originates in \x, and we veto the trigger.  
On the other hand, if the two noise vectors are \emph{inconsistent} with the transfer function, 
we conclude that this particular noise transient in \h\ does not have its origin in \x. In the following 
subsections, we construct two different statistics which can be used to make this decision. One 
statistic is based on constructing a \emph{null-stream} between channel \h\ and the `transferred' channel 
\x; \ie, if the noise transient originates in subsystem \x, and is sensed by channel \x, 
it is possible to construct a linear 
combination of the two data streams such that it does not contain any excess power. The second 
statistic is based on the cross-correlation of the noise in channel \h\ with the `transferred' 
noise in channel \x. 

In the following two subsections, we assume that the data streams $x_i$ and $h_i$ are drawn
from zero-mean Gaussian distributions. Also, we assume that the transfer function from \x\ to 
\h\ is accurately measured/calculated and is time-invariant. In the third subsection, we discuss the 
`real-life scenario' where the data streams are not perfectly Gaussian and the transfer function
is non-stationary. 

\begin{figure}[tb]
\centering
\includegraphics[width=2.2in]{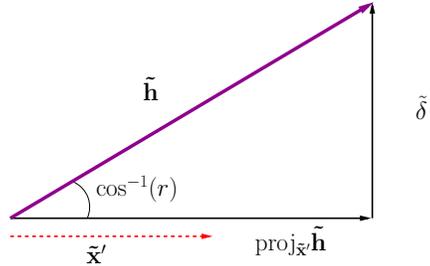}
\caption{Schematic diagram of the linear vector space in which the analysis methods are
formulated.}
\label{fig:VectSpace}
\end{figure}

The linear vector space in which the analysis methods are formulated is schematically
illustrated in Fig.~\ref{fig:VectSpace}.

\subsection{Using the null-stream}
\label{sec:nullStream}
The \emph{null-stream} between $\bth$ and $\btx'$ is the component of the vector $\bth$ orthogonal to
$\btx'$. This can be constructed using the Gram-Schmidt orthogonalisation~\cite{Arfken}: 
\be
\btdelta = \bth - \mathrm{proj}\,_{\btx'} \bth \, ,
\label{eq:delta}
\ee
where we define the projection operator by
\be
\mathrm{proj}\,_{\bf \tilde u}{\bf \tilde v}= \frac{\left<{\bf \tilde v},{\bf \tilde u}\right>}
{\left<{\bf \tilde u},{\bf \tilde u}\right>}\,{\bf \tilde u} \,.
\ee
In the above expression $\left<{\bf \tilde v},{\bf \tilde u}\right>$ denotes the inner 
product between the vectors $\bf \tilde v$ and $\bf \tilde u$: 
\be
\left<{\bf \tilde v},{\bf \tilde u}\right> = {\rm Re} \sum_{k} \tilde v_k \, \tilde u_k^*.
\ee
If the noise transient originates in \x, it will completely disappear 
in $\btdelta$. In order to test this, we compute the \emph{excess power} statistic~\cite{ExPower}
from $\btdelta$:
\be
\epsilon_\delta = \sum_{k}^{} \frac{|\tilde \delta_k|^2}{\sigma_k^2},
\label{eq:stat}
\ee
where $\sigma_k^2$ is the expected variance of $\tilde \delta_k$ in the absence of any excess power. 
In the absence of any excess power in $\btdelta$, $\epsilon_\delta$ will follow a Gamma 
distribution~\cite{JL}. The scale parameter $\alpha$ and shape parameter $\beta$ can be 
estimated from the stationary noise (see~\cite{NPVeto} for more details). 

If $\epsilon_\delta$ is less than, or equal to, a threshold $\tau$, we veto the 
trigger. The threshold $\tau$ giving a rejection probability $\Phi$ (probability
that a `causal' trigger is vetoed) can be calculated from 
\begin{equation}
\Phi = \int_0^\tau \Gamma(x;\alpha,\beta) \, \dx,
\label{eq:exPowThresh}
\end{equation}
where $\Gamma(x;\alpha,\beta)$ is the probability density of the Gamma distribution with 
parameters $\alpha$ and $\beta$.

\subsection{Using the cross-correlation}
\label{sec:crossCorr}
The linear cross-correlation coefficient between two vectors $\btx'$ and $\bth$ is the cosine
of the angle between them: 
\be
 r = {\rm Re} \, \frac{\left< \btx', \bth \right> }{||\btx'||~||\bth||}\,,
\ee
where $||{\bf u}||$ denotes magnitude of the vector ${\bf u}$. If the noise transient
in channel \h\ indeed originates in \x, $\btx'$ and $\bth$ should display a high
correlation. On the other hand,  if the noise transient does not originate in \x, the 
vector $\btx'$ and $\bth$ will be randomly oriented, and hence the linear cross-correlation 
coefficient $r$  will tend to be small in absolute value. This can be converted to 
the normally distributed variable $z$ by the Fisher transformation~\cite{FisherTransform}: 
\be
z = \frac{1}{2} \, {\rm ln} \left( \frac{1+r}{1-r} \right).
\ee
The new variable $z$ will be approximately normally distributed with mean zero and standard 
deviation $1/\sqrt{N-3}$,  where $N$ is the dimension of the vectors $\btx'$ and $\bth$. 

If $z$ is greater than, or equal to, a threshold $\lambda$, we veto the trigger. The 
threshold giving an accidental veto probability of $\psi$ can be calculated from 
\be
\psi = \int_{\lambda}^\infty f(x;\mu,\sigma^2) \, \dx,
\label{eq:crossCorrThresh}
\ee
where $f(x;\mu,\sigma^2)$ is the probability density of the normal distribution with 
mean $\mu = 0$ and variance $\sigma^2 = 1/(N-3)$.

\subsection{Implementation}
\label{sec:realLifeSit}

\begin{figure}[tb]
\centering
\includegraphics[width=2.2in]{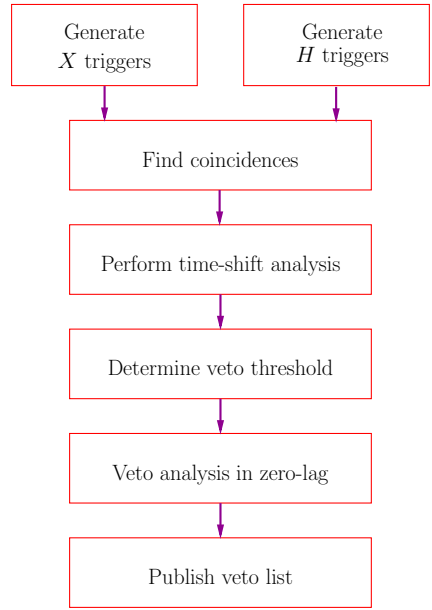}
\caption{Schematic illustration of the veto pipeline.}
\label{fig:VetoPipeline}
\end{figure}

The assumption we made in the previous subsections that the transfer function is time 
invariant is strictly not true. Transfer functions in actual detectors can vary in time. 
The slow temporal variation of the transfer function can be taken into account by making 
repeated measurements of the transfer function and tracking its evolution by continuously 
injecting and measuring spectral lines at certain frequencies (see~\cite{NoiseProj}). 
But the non-stationarities of the transfer function on short time scales are hard to 
track. It may also be noted that the noise in the present-generation interferometers 
is not stationary Gaussian and exhibits tails in the distribution. Considering these 
`real-life' effects, it may not be wise to use the `ideal-case' relations given by 
Eqs.(\ref{eq:exPowThresh}) and (\ref{eq:crossCorrThresh}) to compute the veto thresholds. 
For instance, due to the 
imperfect transfer function, the `null-stream' $\btdelta$ can contain some
`residual burst', and, as a result, the excess power statistic computed from $\btdelta$ will not 
fall into the expected Gamma distribution. But, we do expect the excess power statistic 
$\epsilon_\delta$ computed from $\btdelta$ to be smaller than the same ($\epsilon_{\rm h}$) 
computed from $\bth$. If the ratio $s \equiv \epsilon_{\rm h}/\epsilon_\delta$ is greater 
than a threshold, we veto the trigger. The veto threshold corresponding to a certain 
accidental veto probability is calculated as described below.   

We time shift $x_i$ with respect to $h_i$ to destroy the causal relationship between the 
two data streams~\footnote{Time shift analysis is commonly employed in burst searches
in order to estimate the accidental consistency, or `background' rate. See, for example,~\cite{LSCS4Burst}.}. 
The coincident triggers in the time-shifted data streams are identified
and the `excess-power ratio', $s$, for each coincident trigger is calculated. $n$ such 
time shifts are performed to get better statistics from the data. A threshold, $\tau_s$, 
is chosen such that only an acceptable number of coincident triggers in the time-shifted 
analysis have $s\geq\tau_s$. This threshold $\tau_s$ is used to veto the triggers in 
the `zero-lag' analysis (without time shifting the data). The time-shifted analysis can
also be used to calculate the veto threshold $\lambda$ for the analysis using the 
cross-correlation statistic. Here, we choose a threshold $\lambda$ such that only an 
acceptable number of coincident triggers have $z\geq\lambda$ in the time-shifted analysis
and use this threshold to do the zero-lag analysis. The veto pipeline is schematically
illustrated in Fig.~\ref{fig:VetoPipeline}. 

\subsection{Caveats}
\label{sec:caveats}

It is worth stressing that this method relies on the linearity in the coupling 
of the noise from the detector subsystem \x\ to the detector output, and can not be used 
where the coupling is nonlinear. This method also assumes that the transfer function 
from \x\ to \h\ is unique, and channel \x\ accurately senses the disturbances in \x. 
In other words, this technique can be only applied to systems that exhibit a linear coupling 
through a set path, or multiple paths that are fixed. An environmental monitor will 
often fail to meet this requirement, unless the sensors are exceptionally well placed, 
because each local disturbance 
could couple differently into the monitor and the GW channel, meaning that a different 
transfer function would be needed for each physical point of origin for the disturbance.


\section{Analysis on hardware-injected burst signals}
\label{sec:HWinj}

\subsection{Injections mimicking instrumental bursts}
\label{sec:HWinjInstrBurst}

\begin{figure}[t]
\centering
\includegraphics[width=3.3in]{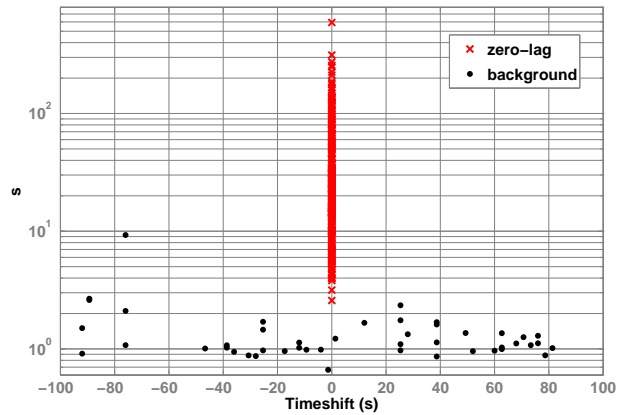}
\caption{Time shift analysis on `instrumental-glitch-like' hardware injections performed 
in the \miphase\ channel.
The horizontal axis shows the time shift applied between $x_i$ and $h_i$, and the vertical 
axis shows the excess-power ratio $s$. The black dots 
correspond to the coincident triggers in the time-shifted analysis and the red crosses 
correspond to the ones in the zero-lag analysis.}
\label{fig:timeShiftExPowHWInj}
\end{figure}

\begin{figure}[t]
\centering
\includegraphics[width=3.3in]{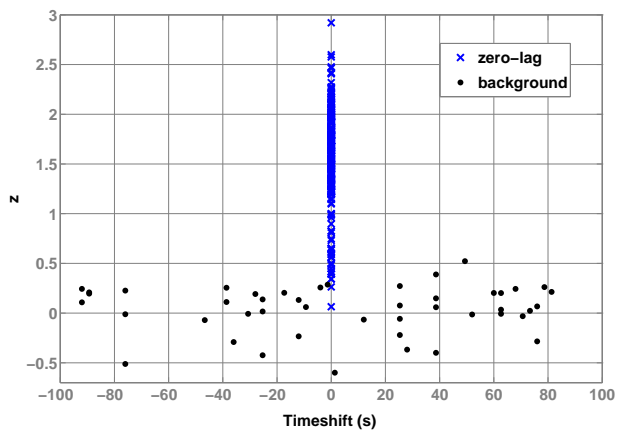}
\caption{Same as in Fig.~\ref{fig:timeShiftExPowHWInj}, except that the vertical 
axis shows the cross-correlation statistic $z$. The black dots correspond to the 
coincident triggers in the time-shifted analysis and the blue crosses correspond to 
the ones in the zero-lag analysis.}
\label{fig:timeShiftCrossCorrHWInj}
\end{figure}

In order to test this veto method, we injected around 300 sine-Gaussian burst signals 
over a period of one hour into four subsystems of \geo, whose couplings to \h\ were known and well 
understood; the injections were performed serially, one subsystem after another. The 
four subsystems we chose are listed below. These descriptions are technical and concise. 
For more information, refer to~\cite{jrSmithThesis}.  

\smallskip
\noindent \textit{Laser amplitude noise} (\micvis): We make bursts of laser amplitude 
noise by injecting glitches into the laser amplitude stabilisation loop. We detect 
these glitches by measuring the light power reflected from the Power-recycling cavity 
in the data acquisition system as channel \micvis.

\smallskip
\noindent \textit{Laser frequency noise} (\micep): We make bursts of laser frequency 
noise by adding glitches to the error-point of the Michelson common-mode control servo 
which keeps the Power-recycling resonant by adjusting the frequency of the master laser. 
The recording of this error-point in the data acquisition system serves as the veto 
channel, \micep.

\smallskip 
\noindent \textit{Michelson oscillator phase noise} (\miphase): The Michelson 
differential arm-length in \geo\ is controlled by imposing phase-modulation
side-bands on the light entering the interferometer. A heterodyne readout scheme is then used to derive
an error signal which is fed back to the end mirrors of the Michelson to keep it on a dark fringe. We
make glitches in the phase of the oscillator used to add the modulation sidebands by driving the
voltage-frequency-control input of the crystal oscillator used to create this modulation signal. We
phase-lock a reference crystal oscillator to the main crystal oscillator and the error-point of the
phase-locked loop, which is sensitive to phase fluctuations on both oscillator signals, is
recorded in the data acquisition system as \miphase\ and serves as a sensitive measurement of the phase
noise on the main Michelson modulation sidebands.

\smallskip
\noindent \textit{Michelson oscillator amplitude noise} (\miamp): The amplitude of 
the the main crystal oscillator is also stabilised to a quiet DC reference. We can 
add signals to the error-point of this stabilisation servo so as to impose additional 
amplitude noise on the main Michelson modulation signal. We added glitch signals
in to this control loop and recorded its error-point in the data acquisition system 
as \miamp\ to serve as a veto channel.

\smallskip
Burst triggers in the veto channel and the GW channel are generated using the mHACR~\cite{Hild07,NPVeto}
burst detection algorithm. mHACR belongs to the class of time-frequency detection algorithms those 
make a time-frequency map of the data and identify time-frequency pixels containing excess 
power which are statistically unlikely to be associated with the underlying noise 
distribution. For a detailed description of the algorithm and its performance, see ~\cite{Hild07}. 
However, we remind the reader that the details of the burst detection algorithm are 
immaterial as far as this veto method is concerned. The burst ETG is only used to identify
the coincident triggers in the two channels, and any ETG with proper time estimation of
the burst event should serve this purpose. 

Coincident triggers in the two channels are identified using a  time window of $\pm10$\,ms. 
The results of the time shift analysis on the hardware injections performed in the \miphase\ are shown in
Figs. \ref{fig:timeShiftExPowHWInj} and \ref{fig:timeShiftCrossCorrHWInj}. The horizontal axis shows the
time shift applied between $x_i$ and $h_i$. The vertical axis in Fig. \ref{fig:timeShiftExPowHWInj}
shows the excess-power ratio $s \equiv \epsilon_{\rm h}/\epsilon_\delta$. The (black) dots correspond to the
coincident triggers in the time-shifted analysis and the (red) crosses correspond to the ones in the
zero-lag analysis. From this, a veto threshold of 2.35 is chosen which corresponds to an accidental
veto rate of 1 per day. All the coincident triggers in the zero lag are vetoed using this threshold. The
vertical axis in Fig. \ref{fig:timeShiftCrossCorrHWInj} shows the (transformed) cross-correlation
statistic $z$. The (black) dots correspond to the coincident triggers in the time-shifted analysis and the
(blue) crosses correspond to the ones in the zero-lag analysis. The veto threshold corresponding to an
accidental veto rate of 1 per day is 0.27, which resulted in vetoing 99\% of the coincident triggers in the
zero-lag.

The veto analysis is performed on all the four channels in which the hardware injections
are done. Results of the analysis are summarised in Table \ref{table:HWresultsSummary}. 
It can be seen that only $\sim 5\%$ of the coincident triggers in the time-shifted
analysis are vetoed, while more than 90\% of the coincident triggers in the zero-lag are 
vetoed. This implies that the accidental rate of the veto is only $\sim 5\%$ of that of the standard 
statistical veto (using a time window of $\pm10$\,ms) for almost the same veto efficiencies.  

\begin{table}[htbp]
    \begin{center}
        \begin{tabular}{cccccccccc}
            \hline
            \hline
            Veto &\vline& \multicolumn{2}{c}{Threshold} &\vline& \multicolumn{5}{c}{Veto fraction} \\
            \cline{2-10}
            channel  &\vline& $\tau_s$ & $\lambda$ &\vline& $\xi $ && $\chi_s$ && $\chi_z$ \\
            \hline
            \micep   &\vline&  2.51 & 0.33 &\vline& $4.48 \times 10^{-2}$ && 0.90 && 0.90 \\
            \micvis  &\vline&  1.94 & 0.23 &\vline& $5.45 \times 10^{-2}$ && 1.00 && 1.00 \\
            \miphase &\vline&  2.35 & 0.27 &\vline& $6.12 \times 10^{-2}$ && 1.00 && 0.99 \\
            \miamp   &\vline&  1.50 & 0.26 &\vline& $4.62 \times 10^{-2}$ && 0.97 && 0.97 \\
            \hline
            \hline
        \end{tabular}
        \caption{Summary of the veto analysis on hardware injections mimicking instrumental 
        bursts. $\tau_s$ and $\lambda$ are the chosen veto thresholds on the excess-power ratio $s$ 
        and the cross-correlation statistic $z$, respectively. 
        $\xi $ is the fraction of \emph{coincident} events that are vetoed in the time-shifted 
        analysis. The fraction of coincident events vetoed in the zero-lag using
        the $s$ statistic is denoted by  $\chi_s$, while the same using the $z$ statistic is 
        denoted by  $\chi_z$. The chosen thresholds correspond to an accidental veto rate of 1 per day.}
        \label{table:HWresultsSummary}
    \end{center}
\end{table}
%
\subsection{Injections mimicking gravitational-wave bursts}
\label{sec:HWinjGWBurst}

Some of the interferometer channels are sensitive to GWs to some non-negligible level. This raises the
question of \emph{veto safety} while using interferometer channels as veto channels. \ie, we have to
make sure that we do not veto actual GW bursts which are coincident in the two channels. We argue that,
since actual GW bursts are not causally related to the instrumental channels, the coincident triggers in
channels \x\ and \h\ will not be consistent with the transfer function from \x\ to \h\, and hence, will
not be vetoed using this method. Although the four channels under investigation in this paper are not
expected to show any non-negligible sensitivity to GWs, there can be unexpected couplings, for example,
through electrical faults or cross couplings in the data acquisition system. It is therefore prudent to
explicitly demonstrate the safety of this veto method by doing GW-like hardware injections.

Hardware injections are performed by injecting signals into the electrostatic actuators used to control
the differential-arm-length degree of freedom of \geo. For the test described here, around 300
sine-Gaussian bursts were injected with varying amplitudes and with central frequencies in the range 200
to 1300\,Hz.

\begin{figure}[t]
\centering
\includegraphics[width=3.3in]{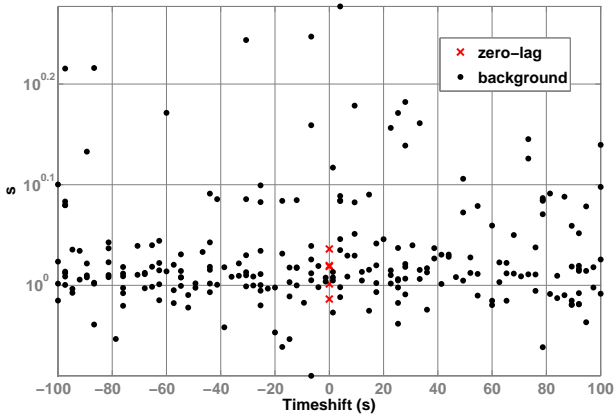}
\caption{Time shift analysis on the GW-like hardware injections. $E_{\rm ref}$ is 
used as the veto channel. The horizontal axis shows the time shift
applied between $x_i$ and $h_i$, and the vertical axis shows the excess-power ratio 
$s$. The black dots correspond to the coincident triggers
in the time-shifted analysis and the red crosses correspond to the ones in the zero-lag
analysis.}
\label{fig:timeShiftExPowGWLikeHWInj}
\end{figure}

\begin{figure}[t]
\centering
\includegraphics[width=3.3in]{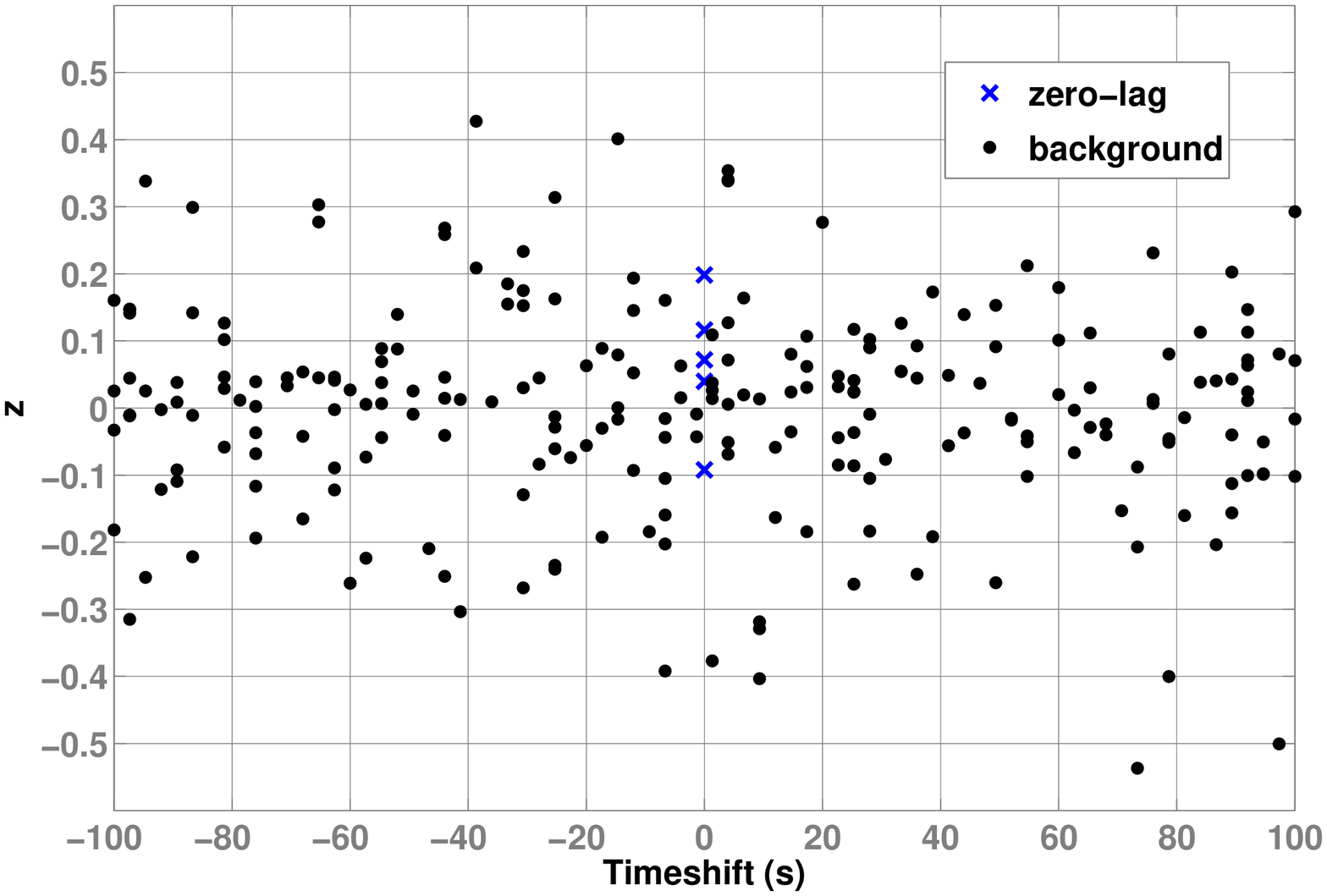}
\caption{Same as in Fig.~\ref{fig:timeShiftExPowGWLikeHWInj}, except that the 
vertical axis shows the cross-correlation statistic
$z$. The black dots correspond to the coincident triggers in the time-shifted analysis 
and the blue crosses correspond to the ones in the zero-lag analysis.}
\label{fig:timeShiftCrossCorrGWLikeHWInj}
\end{figure}

Figs.~\ref{fig:timeShiftExPowGWLikeHWInj} and \ref{fig:timeShiftCrossCorrGWLikeHWInj}
show the results of the veto analysis performed on the GW-like hardware injections. 
\micep\ is used as the veto channel. It can be seen that neither of the test statistics 
($s$ or $z$) in the zero-lag analysis shows any excess significance over the corresponding 
time-shifted analysis. The veto thresholds corresponding to an accidental rate of 1 per 
day are $\tau_{\rm s} = 1.64$ and $\lambda = 0.35$. Using these thresholds, we don't 
veto any of the injections. 

\section{An example application}
\label{sec:applRealData}

\begin{figure}[t]
\centering
\includegraphics[width=3.3in]{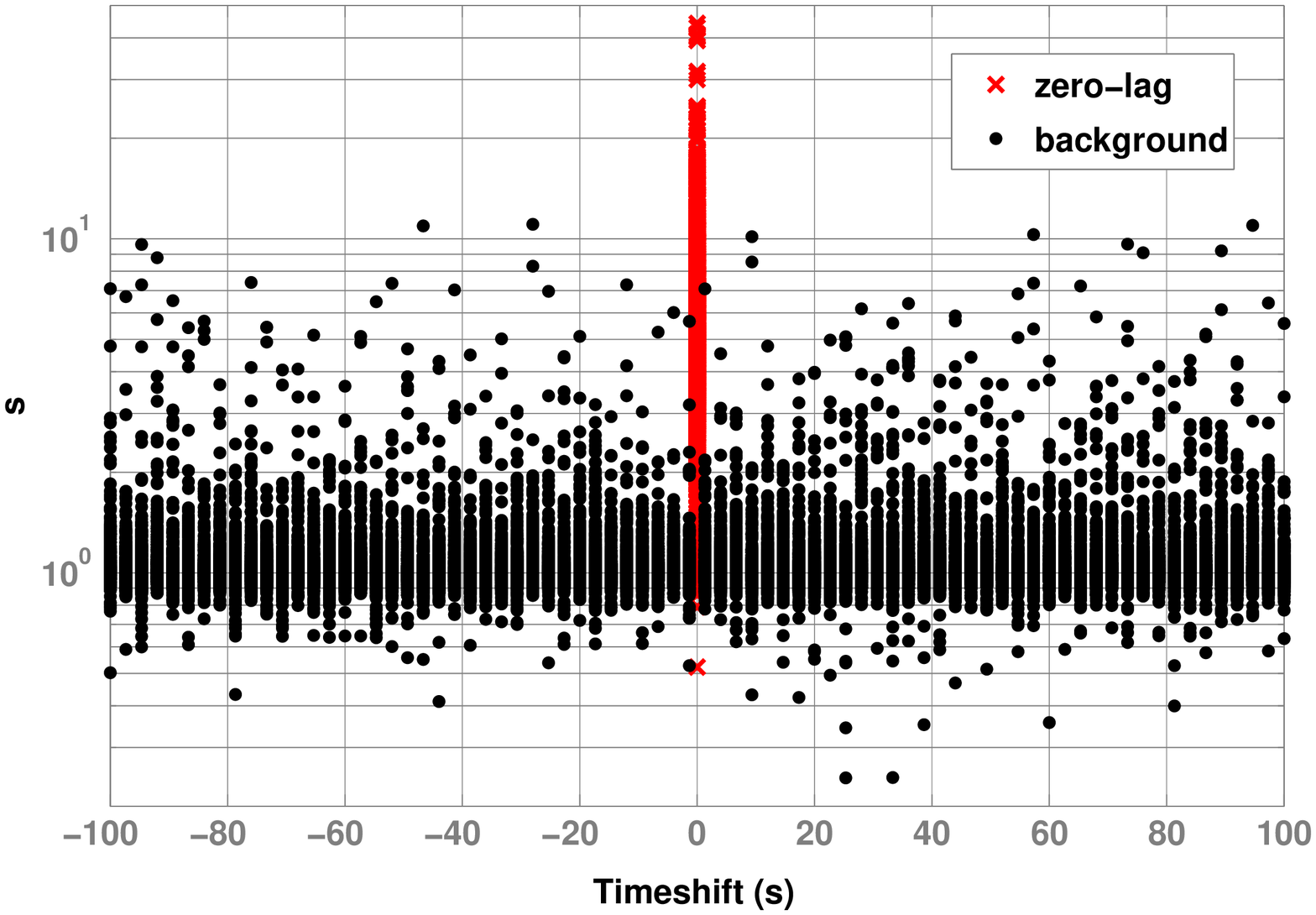}
\caption{Time-shifted analysis on 5 days of data from the Fifth science run of \geo\
using \micep\ as the veto channel. The horizontal axis shows the time shift
applied between $x_i$ and $h_i$, and the vertical axis shows the excess-power ratio 
$s$. The black dots correspond to the coincident triggers
in the time-shifted analysis and the red crosses correspond to the ones in the zero-lag
analysis.}
\label{fig:S5MICEPExPowTimeShiftPlot}
\end{figure}

\begin{figure}[t]
\centering
\includegraphics[width=3.3in]{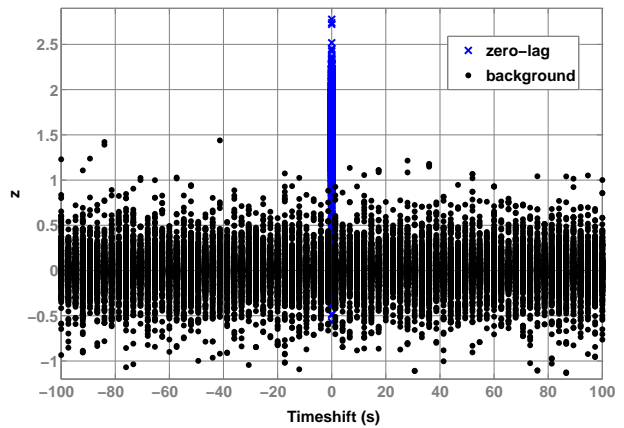}
\caption{Same as in Fig.~\ref{fig:S5MICEPExPowTimeShiftPlot}, except that the vertical 
axis shows the cross-correlation statistic
$z$. The black dots correspond to the coincident triggers in the time-shifted analysis 
and the blue crosses correspond to the ones in the zero-lag analysis.}
\label{fig:S5MICEPCrossCorrTimeShiftPlot}
\end{figure}

\geo\ participated full time in the `fifth science run' (S5 run) in coincidence with the LIGO detectors
from May 2006 to October 2006. The first few weeks of \h\ data contained an additional population of
glitches identified as coming from the laser frequency stabilisation control loop. These excess glitches
had central frequencies typically around 2\,kHz. In fact, the glitches were broad-band in the frequency
stabilisation loop, but the coupling of frequency noise to \h\ is most prominent around 2\,kHz and so
this is where we see the excess noise in \h. 
The identification and repair of the source of these glitches took several weeks.
So we identified an appropriate measure (\micep) of the
frequency-noise glitches that could be used as a veto channel (see Sec.~\ref{sec:HWinjInstrBurst} to see
how this channel is derived).

Veto analysis is performed on 5 days of data from the period described above (in the 
frequency range of 400 Hz -- 2kHz). Burst triggers in the two channels are generated 
by the mHACR burst-detection algorithm. Coincident triggers within the two channels 
are identified using a time window of $\pm10$ ms for time-coincidence. 
Only triggers with signal-to-noise ratio $\geq\,6$ are considered for this analysis. Out of 5326 
triggers in the GW channel, 2048 triggers were found to be coincident with the \micep\ channel. 
The accidental rate of the veto is estimated by doing 76 time shifts (from -100s to 100s). 
Fig.~\ref{fig:S5MICEPExPowTimeShiftPlot} shows the excess-power ratio $s$ computed from 
the coincident triggers plotted against the applied time shift between the data streams,
while Fig.~\ref{fig:S5MICEPCrossCorrTimeShiftPlot} shows the cross-correlation statistic
$z$ plotted against the time shift. the dots correspond to the coincident triggers from
the time-shifted analysis and the crosses correspond to those from the zero-lag analysis. 
We choose an accidental veto rate of 1 per day. The thresholds
on the two statistics are estimated from the time-shifted analysis. This corresponds to 
a threshold of $\tau_{\rm s} = 2.25$ for the excess-power ratio $s$ and a threshold of 
$\lambda = 0.54$ for the cross-correlation statistic $z$. In the analysis using the null-stream,
all coincident triggers with $s \geq \tau_{\rm s}$ are vetoed, while in the analysis
using the cross-correlation statistic, all triggers with $z \geq \lambda$ are vetoed.
It was found that 88\% of the coincident triggers are vetoed using the null-stream method and 92\% of 
the coincident triggers are vetoed using the cross-correlation method. These correspond to 34\% 
and 35\% of the total number of \h\ triggers in the data.  

\begin{figure}[t]
\centering
\includegraphics[width=3.3in]{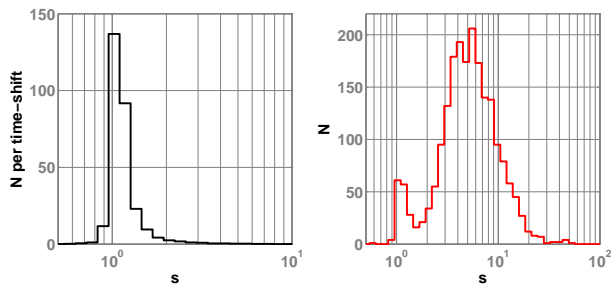}
\caption{Histograms of the excess-power ratio $s$ computed 
from the time-shifted analysis (left) and the zero-lag analysis (right).}
\label{fig:MICEPS5JuneExPowHist}
\end{figure}

\begin{figure}[t]
\centering
\includegraphics[width=3.3in]{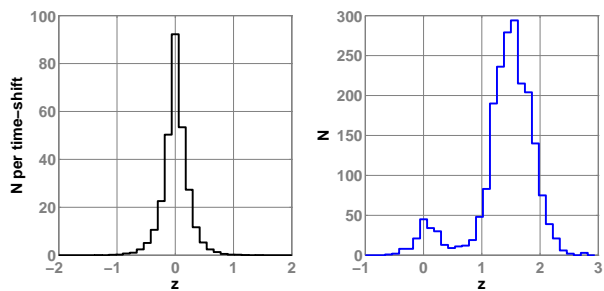}
\caption{Histograms of the cross-correlation statistic $z$ computed from the 
time-shifted analysis (left) and the zero-lag analysis (right).}
\label{fig:MICEPS5JuneCrossCorrHist}
\end{figure}

\begin{figure}[t]
\centering
\includegraphics[width=3.3in]{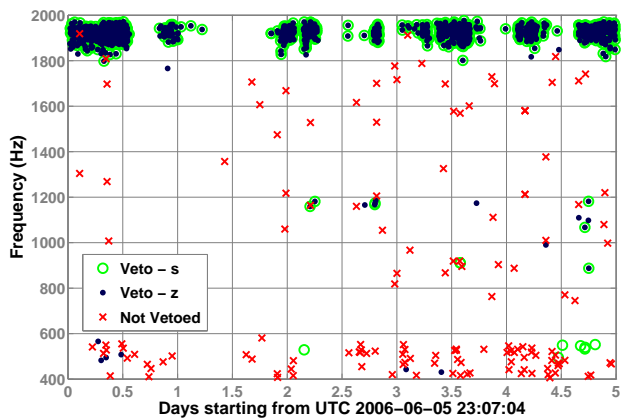}
\caption{A time-frequency plot of mHACR triggers from 5 days of \geo\ data. The horizontal
axis shows the time and the vertical axis shows the frequency of the 
the burst triggers in channel \h, as estimated by mHACR. Only those triggers which are
coincident with \micep\ are plotted. The (green) circles correspond to coincident triggers which are 
vetoed using the null-stream method, and the (black) dots correspond to the ones which 
are vetoed using the cross-correlation method. Coincident triggers which are not vetoed
by any of the methods are indicated by (red) crosses. The chosen veto thresholds correspond
to one accidental veto per day.}
\label{fig:HACRTFplotVeto}
\end{figure}

Histograms of the two test statistics $s$ and $z$ computed from the coincident triggers 
are plotted in Figs.~\ref{fig:MICEPS5JuneExPowHist} and \ref{fig:MICEPS5JuneCrossCorrHist}. 
The plots on the left show the distributions of the test statistics computed from
the time-shifted analysis, normalised by the number of time shifts applied. These are 
the expected distributions of $s$ and $z$ in the absence of any causal relation between triggers
in \x\ and \h\ (for the given data set). Histograms on the right show the distributions
of $s$ and $z$ computed from the zero-lag analysis. Two different populations are 
clearly visible in these plots. One population (centered around 1 in the histograms of $s$; 
centered around zero in the histograms of $z$) corresponds to the triggers which are
accidentally time-coincident in the channels \x\ and \h, while the other population
(centered around 6 in the histogram of $s$; centered around 1.6 in the histogram of $z$)
corresponds to triggers in \h\ which are causally related to the ones in \x. It is 
interesting to note that this method is able to distinguish clearly between these two 
populations.
The reader may note that the number of accidental coincidences in the time-shifted 
analysis is $\sim 35\%$ larger than that in the zero-lag analysis. 
This can be explained in the following way: the veto analysis has shown 
that around 35\% of the triggers in channel \h\ are causally related to \micep. 
These `causal' triggers will fall into the populations on the right in the zero-lag analysis
(centered around 6 in the histograms of $s$; centered around 1.6 in the histograms of $z$).
But, since the accidental coincidence rate is directly proportional to the total number of triggers, 
the presence of these causal triggers in the data would increase the coincidence rate in 
the time-shifted analysis by $\sim 35\%$, thus explaining the excess-coincidences that
we observe.

Fig.~\ref{fig:HACRTFplotVeto} shows a time-frequency plot of the mHACR triggers from
5 days of \geo\ data. The green circles correspond to the coincident triggers (in channel \h)
which are vetoed using the null-stream method and the black dots correspond to the ones 
which are vetoed using the cross-correlation method. The red crosses correspond to the 
coincident triggers which are not vetoed by any of the methods. 

\begin{figure}[t]
\centering
\includegraphics[width=3.1in]{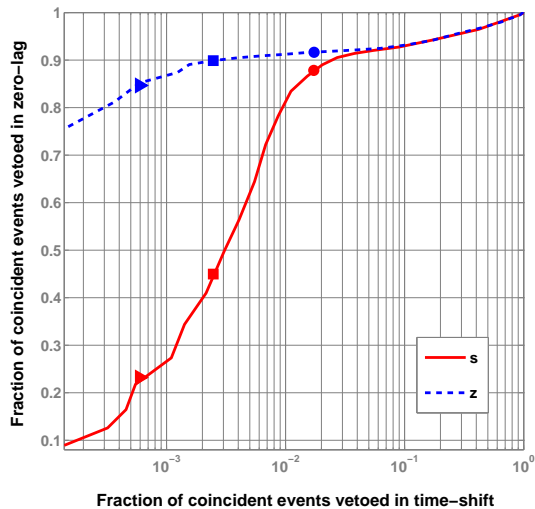}
\caption{Fraction of coincident triggers that are vetoed in the zero-lag plotted against 
the fraction of the coincident triggers that are vetoed in the time-shifted analysis. 
The solid curve corresponds to the analysis using null-stream and the dashed curve
corresponds to the analysis using cross-correlation. The triangles, the squares, and
the dots correspond to accidental veto rates of 1 per month, 1 per week, and 1 per day, 
respectively.}
\label{fig:ROCplot}
\end{figure}

A summary of the analyses performed using different accidental veto rates are given in Table
\ref{table:S5resultsSummary}. Also, in Fig.~\ref{fig:ROCplot}, we plot the fraction 
of coincident events that are vetoed in the zero-lag (a measure of the efficiency of 
the veto) against the fraction of the coincident triggers which are vetoed in the
time-shifted analysis (a measure of the accidental veto probability). This plot can 
be thought of as a \emph{receiver operating characteristic}~\cite{LouisScharf} plot for this
analysis, and can be used to choose thresholds which correspond to acceptable 
values of veto efficiency and accidental veto rate/probability. In the Figure, 
the solid curve corresponds to the analysis using null-stream and the dashed curve
corresponds to the analysis using cross-correlation.  
It can be seen that, for high values of accidental veto probability, the two methods
perform equally well. But for low values of accidental veto probability ($< 2\times10^{-2}$),
the curve corresponding to the null-stream analysis starts to fall off, and the 
cross-correlation analysis continues to perform well.  

\begin{table}[htbp]
    \begin{center}
        \begin{tabular}{cccccccccc}
            \hline
            \hline
            Accidental &\vline& \multicolumn{2}{c}{Threshold} &\vline& \multicolumn{5}{c}{Veto fraction} \\
            \cline{2-10}
            rate  &\vline& $\tau_s$ & $\lambda$ &\vline& $\xi $ && $\chi_s$ && $\chi_z$ \\
            \hline
            ${\rm day}^{-1}$   &\vline& 2.25  & 0.54 &\vline& $1.73 \times 10^{-2}$ && 0.88 && 0.92 \\
            ${\rm week}^{-1}$  &\vline& 5.09  & 0.86 &\vline& $2.47 \times 10^{-3}$ && 0.45 && 0.90 \\
            ${\rm month}^{-1}$ &\vline& 7.40  & 1.11 &\vline& $5.93 \times 10^{-4}$ && 0.23 && 0.85 \\
            \hline
            \hline
        \end{tabular}
        \caption{Summary of the veto analysis on 5 days of data from \geo. 
        $\tau_s$ and $\lambda$ are the chosen veto thresholds on the excess-power ratio $s$ 
        and the cross-correlation statistic $z$, respectively. $\xi $ is the fraction of 
        \emph{coincident} events that are vetoed in the time-shifted analysis. The fraction 
        of coincident events vetoed in the zero-lag using the $s$ statistic is denoted by  
        $\chi_s$, while the same using the $z$ statistic is denoted by  $\chi_z$.
        The chosen veto thresholds correspond to the accidental veto rates tabulated in
        the first column. }
        \label{table:S5resultsSummary}
    \end{center}
\end{table}

\begin{figure}[t]
\centering
\includegraphics[width=3.0in]{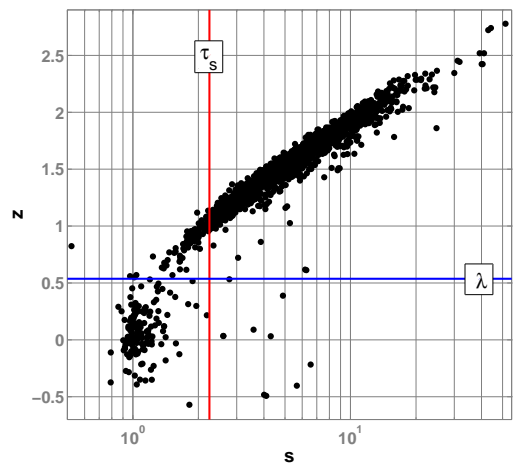}
\caption{Excess-power ratio $s$ (horizontal axis) computed from the coincident triggers
plotted against the cross-correlation statistic $z$ (vertical axis). The vertical (red)
and  horizontal (blue) lines correspond to the veto thresholds $\tau_s$ and $\lambda$ on the 
two statistics, respectively.}
\label{fig:exPowVsCrossCorr}
\end{figure}

Fig.~\ref{fig:exPowVsCrossCorr} provides a rough comparison between the abilities of the
two test statistics ($s$ and $z$) in vetoing the instrumental glitches. The horizontal 
axis shows the  excess-power ratio $s$ and the vertical axis shows  
the cross-correlation statistic $z$ computed from the coincident triggers. The vertical (red)
and  horizontal (blue) lines in the plot correspond to the veto thresholds $\tau_s$ and
$\lambda$ on the two statistics, respectively. Triggers on the right of the vertical line are vetoed
by $s$, and those above the horizontal line are vetoed by $z$. 33.5\% of the total number
of \h\ triggers are vetoed by \emph{both} methods. There exists a small population 
($\sim 1\%$ of the total number of \h\ triggers) which is vetoed by $z$; but not by $s$, 
which suggests that $z$ is a more sensitive statistic than $s$. But this may 
not be taken as a general indication that the cross-correlation is a more sensitive
method than the null-stream. One can construct alternative statistics using the null-stream, 
which could be more sensitive than $s$. One possible alternative is
$\epsilon_\delta/\epsilon_{\rm opt}$, where $\epsilon_{\rm opt}$ is the excess power statistic
computed from the optimal combination~\cite{LazzariniOptH,HewitsonOptH} of the noise vectors 
$\bth$ and $\btx'$.


\section{Summary and outlook}
\label{sec:summary}

One of the most challenging problems in the search for unmodelled GW bursts using 
ground-based detectors is to distinguish between actual GW bursts and spurious instrumental 
bursts that trigger the detection algorithms. In this paper, we proposed a veto 
method which makes use of the information on the physical coupling of different 
detector subsystems to the main detector output. We also demonstrated this method 
using the data of the \geo\ detector. By performing hardware injections mimicking instrumental 
glitches, we showed that glitches originating in a detector subsystem can be 
vetoed using the transfer function from the subsystem to the detector output. We 
also addressed the issue of veto safety by performing hardware injections mimicking 
GW bursts into \geo, and by showing that such injections are \emph{not} vetoed. Finally, 
we used this strategy to veto glitches in the data from the fifth science run of 
\geo, using the laser frequency noise channel as the veto channel. 
The analysis was performed on 5 days of \geo\ data from the second month of the  
science run. 35\% of the 5326 triggers in the GW channel were vetoed with an accidental 
rate of 1 per day using the `cross-correlation' method, while 34\% of the triggers 
were vetoed using the `null-stream' method. 

The method relies on linearity in the coupling of the noise from a detector subsystem 
to the detector output, and the measurability/calculability and uniqueness of the 
transfer function. The assumption of linear coupling is valid 
as far as many detector subsystems in the large-scale interferometers are concerned. 
Strictly speaking, this method also requires time-invariant transfer functions. The way
to track down slow temporal variations in the transfer functions is discussed in
the paper. The formulation that we have developed was found to be robust against non-stationarities 
of short time scales in the transfer functions, and non-Gaussian tails in the noise
distribution. 

When possible, using physical information has clear advantages over relying only 
on statistical correlations. The method that was proposed in this paper is 
a fully coherent way of testing the consistency of the glitches in the GW channel
with those in an instrumental channel. The authors hope that this will serve as a first step for 
developing a class of `physical instrumental vetoes' for present and future detectors.

\section*{Acknowledgments}

The authors are grateful for support from PPARC and the University of Glasgow 
in the UK, and the BMBF and the state of Lower Saxony in Germany.
The authors also thank Peter Shawhan and Peter Saulson for their detailed
comments on the manuscript, and the members of the \geo\ group of the 
Albert Einstein Institute for useful discussions. This document has 
been assigned LIGO Laboratory document number LIGO-P070032-00-Z. 

\bigskip


\end {document}